# Micro-Tesla Offset in Thermally Stable AlGaN/GaN 2DEG Hall-effect Plates using Current Spinning

Karen M. Dowling[1], Hannah S. Alpert[2], Ananth Saran Yalamarthy[3], Peter F. Satterthwaite[1],

Sai Kumar[2], Helmut Köck[4], Udo Ausserlechner[4], Debbie G. Senesky[1,3,a]

[1]*Department of Electrical Engineering, Stanford University, Stanford, California, 94305 USA*
[2]*Department of Aeronautics and Astronautics, Stanford University, Stanford, California, 94305 USA*
[3]*Department of Mechanical Engineering, Stanford University, Stanford, California, 94305 USA*
[4]*Infineon Technologies, AG, Villach, 9500, Austria*

*Abstract*—This letter describes the characterization of a low-offset Hall-effect plate using the AlGaN/GaN two-dimensional electron gas (2DEG). Four-phase current spinning was used to reduce sensor offset voltage to values in the range of ~20 nV, which corresponds to a low residual offset of ~2.6 µT when supplied with low voltages (0.04 V to 0.5 V). These offsets are 50x smaller than the values previously reported for GaN Hall-effect plates, and it is on par with state-of-the-art silicon Hall-effect plates. In addition, the offset does not exceed 10 µT even at higher supply voltage of 2.34 V. The sensor also shows stable current-scaled sensitivity over a wide temperature range of -100ºC to 200ºC, with temperature drift of -125 ppm/ºC. This value is 3x better than state-of-the art silicon Hall-effect plates. Additionally, the sensor's voltage-related sensitivity (~57 mV/V/T) is similar to that of state-of-the-art silicon Hall-effect plates. Because of their low offset values, AlGaN/GaN 2DEG Hall-effect plates are viable candidates for low-field and high temperature magnetic sensing in monolithic GaN systems used in extreme temperature environments such as power inverter, down-well, combustion, and space applications.

a) Author to whom correspondence should be addressed: dsenesky@stanford.edu

**Introduction**

Magnetic field sensing is widely used for both direct purposes such as navigation using the Earth's field and for indirect purposes such as motor position or current monitoring. In most applications, the ideal magnetic field sensor would exhibit a high sensitivity to maximize the output signal, and low offset to accurately detect small fields. While giant magnetoresistance (GMR) and tunnel magnetoresistance (TMR) sensors have the highest sensitivities, they suffer from hysteresis, behave non-linearly, and have large offsets in DC applications [1]. Hall-effect plates have better linearity than such devices and are easily fabricated in integrated circuit (IC) technology. Furthermore, current spinning mitigates large raw offsets in Hall-effect plates without external calibration [3], [6], [7], [13]–[21]. This strategy takes advantage of device symmetry to subtract small nonidealities and create a Hall voltage with near-zero offset. Current spinning was popularized by Munter in 1989 [16] with use of complementary metal-oxide-semiconductor (CMOS) analog technology, and has typically been implemented in devices with bulk-film doping. Recent work shows that this approach can also be used to reduce offsets in two-dimensional electron gas (2DEG) structures, such as those based on GaAs [22], [23]. GaN, a popular 2DEG platform for power electronics, would also benefit from low offset, highly sensitive

Hall-effect plates that can be monolithically integrated with power devices for in-situ current monitoring [10].

In this paper, we close a literature gap in AlGaN/GaN magnetic sensing devices by examining a low offset 2DEG Hall-effect plate that is competitive with silicon (Si) Hall plates, as shown in Table I. In addition, the sensor can be readily monolithically integrated with GaN IC's and power devices. We implemented a 4-phase current spinning technique to significantly reduce sensor offset ($B_0$) within a range of 2.6 to 10 µT with supply voltages from 44 mV to 2.34 V. It should be noted that our results are limited by the measurement equipment to about 10 nV, yet are comparable to the lowest obtained values for Si Hall-effect plates [3], [6], [7] and are 50x smaller than previously published GaN offset values [11], [12]. Our sensor exhibits a voltage-scaled sensitivity ($S_V$) of 57 mV/V/T and current-scaled sensitivity ($S_I$) of 89 V/A/T at 25ºC, comparable to that of previous GaN [8]–[10] and Si Hall-effect plates [2], [5]. Finally, we measured the temperature coefficient of drift (TCD) of $S_I$ to be -125 ppm/°C, which is ~3x smaller than Si Hall-effect plates.

**Experiment**

Devices were fabricated with a metal-organic chemical vapor deposited AlGaN/GaN on <111> Si substrate purchased from DOWA, using a process similar to that of Yalamarthy *et al*., 2018 [24]. We obtained a sheet resistance of 430 Ω/□ and contact resistance of $1 \times 10^{-5}$ Ω-cm² as indicated from transfer length method (TLM) measurements using structures also fabricated on the same wafer. An image of the Hall-effect plate is in Fig. 1a. The device has a 200-µm-diameter octagon with 70-µm-long legs extending from 4 sides for electrical contact. The longitudinal resistance was 1560±3 Ω. The device was then wire-bonded to a printed circuit board (PCB) for testing in a home-made shielded Helmholtz coil pair described in previous work [11]. Another sensor from the same wafer was also tested from -100°C to 200°C in a similar, unshielded test apparatus.

This sensor operates via the Hall-effect [18]; supply current ($I_S$) is applied across the sensor in the presence of a magnetic field and Hall voltage ($V_H$) is measured across the other electrodes. Two particular metrics of sensitivity for Hall-effect plates are $S_I$ and $S_V$. These terms are previously described by Lu *et al*. in 2006 [9] and geometry factors were clarified in later work. It is well known that $S_I$ has an inverse relationship to the sheet density of the 2DEG, and $S_V$ is directly proportional to the electron mobility in the 2DEG. While both terms also have relationships to the geometrical shape of the plate, this is not the focus of this work.

Key sources of Hall-effect offset voltages ($V_0$) are from resistive asymmetry due to device fabrication and material defects, asymmetric self-heating, packaging effects, and external instrument noise [4], [11]. This diverse set of contributions leads to a varied $V_0$ in different measurement configurations. In particular, measurements taken with 90° rotated source and sensor electrodes will have $V_0$ with opposite polarity, as shown in Fig. 1b (Phase A, B, C, and D). Current spinning is possible due to this switching polarity (resulting from static sources of offset such as resistance asymmetry), and one can calculate the true $V_H$ from these 4 unique Hall voltage configurations ($V_{A-D}$). This can be understood by a simple Wheatstone bridge representation of a Hall-effect plate, described elegantly by Bilotti *et al.* in 1997 [13]. Static, linear offsets are removed with 4-phase current spinning using (1),

$$V_H = \frac{V_A + V_B + V_C + V_D}{4} = S * B + V_{0,res.} \qquad \text{Equation 1}$$

When applying this calculation, only the induced voltage from the product of the magnetic field ($B$) and sensitivity ($S$), as well as the residual offset voltage ($V_{0,res}$) from imperfect cancellations between the different measurement phases remain. The current spinning was implemented using the same procedure described previously [11].

**Results and Discussion**

Raw offsets are a major portion of the measured Hall voltage, as shown in Fig. 1b. When we implemented current spinning with Equation 1, only small residual offsets remained (Fig. 1c). Fig. 2a shows the magnitude of the raw offset voltage of the 4 phases measured and the $V_{0,res}$ for varied supply voltage ($V_S$). $V_{0,res}$ varied from 0.8 nV to 600 nV for $V_S$ from 46 mV to 2.34 V. The attenuation of raw offset to residual offset was around 600 to 780, which represents a significant improvement in the minimum detectable magnetic field.

To compare between different sensor technologies, Hall offsets from erroneous voltages are usually discussed in literature as magnetic offsets ($B_{0,res}$). One calculates $B_{0,res}$ from $V_{0,res}$ and $S$. Here, the offset confidence ($C$) was calculated for 300 measurements at each supply condition. When $V_S > 0.3$ V, the $B_{0,res}$ was 5.6 µT with $C < 2$ µT. At $V_S < 0.3$ V the average residual offset remains low, but $C$ was as high as 20 µT, shown in Fig. 2b. The standard deviation of $V_{0,res}$ was 0.63 µV regardless of $V_S$, which implies large $C$ at low $V_S$ is due to instrumentation limits from the switching matrix's thermal emf ($< 3$ µV) and measurement accuracy of the Agilent 34401A multimeter, as well as the background field of the shielding chamber [14]. Additional measurements to determine the baseline were done with a wheatstone bridge with 1 kΩ resistors, and showed a minimum measurement signal of around 10 nV.

$B_{0,res}$ values in current-spun Si Hall-effect plates have been reported as low as 2.5 µT with similar improvements in ratio with raw offset [3], [13], [16], [17], [19]. Since the lowest offsets found here were between 0.3 and 2.7 µT, it is clear that current spinning is beneficial for Hall-effect plate operation in AlGaN/GaN to enable competitive near-zero offset sensors.

The $S_I$ and $S_V$ measured from -100ºC to 200ºC are shown in Fig. 2c. $S_V$ is 57 mV/V/T at 25ºC, with TCD of -0.7 %/ºC since mobility decreases with temperature. The same mobility dependence is observed with sheet resistance which increased from 430 Ω/□ at 25ºC to 1 kΩ/□ at 150ºC. $S_V$ could be improved through optimized design of the Hall-effect plate geometry [25]. In silicon, $S_I$ is usually higher than that of AlGaN/GaN [2], [20], because of lower sheet density than the 2DEG. Here, the AlGaN/GaN $S_I$ is 89 V/A/T at 25ºC. The TCD of $S_I$ of this 2DEG Hall-effect plate is approximately -125 ppm/ºC, consistent with previous reports [8], [9]. As described previously, $S_I$ is inversely related to the 2DEG sheet density, which is affected by the crystals' spontaneous and piezoelectric polarizations (inert in this temperature range [26], [27]). However, the sheet density in Si-based devices is from external ionizing dopants, so their $S_I$ TCD is larger in this temperature range than a 2DEG device based on the AlGaN/GaN material systems. Previous work reported TCD of -800 ppm/ºC and -336 ppm/ºC for $S_I$ in Si Hall-effect plates [2], [28], so there is 3x stability improvement for the AlGaN/GaN Hall-effect plate due to suppression of thermally induced intrinsic carriers. Thus, 2DEG Hall-effect plates could be implemented with constant-current circuits when temperature stability is required.

**Conclusion**

In this letter, we presented a 2DEG GaN Hall-effect plate with record low offset of 2.6 µT (1/20th of Earth's magnetic field), which is on par with the best reported silicon Hall-effect plate offsets. We accomplished this by using a 4-phase current spinning technique to reduce the raw offset in AlGaN/GaN 2DEG Hall-effect plates by 3 orders of magnitude. We also confirmed robust sensor operation at extreme temperatures from -100°C to 200°C, which is beyond the typical operation range of Si devices. This work has shown the promising potential for use of GaN magnetic-field sensors as a wide temperature range sensor solution in monolithically integrated GaN electronics for power systems, autonomous position sensing, and novel space exploration applications.

TABLE I - COMPARISON OF GaN AND Si HALL-EFFECT SENSORS

| Material | $S_I$ (V/A/T) | $S_I$ TCD (ppm/°C) | $S_V$ (mV/V/T) | $B_0$ (µT) | Refs |
|---|---|---|---|---|---|
| Si | 107 | 300-800 | 33-72 | 2.5 | [2]–[7] |
| GaN 2DEG | 55-113 | 102 | 76 | 100 | [8]–[12] |
| GaN 2DEG | 89 | 125 | 57 | 2.6 | This Work |

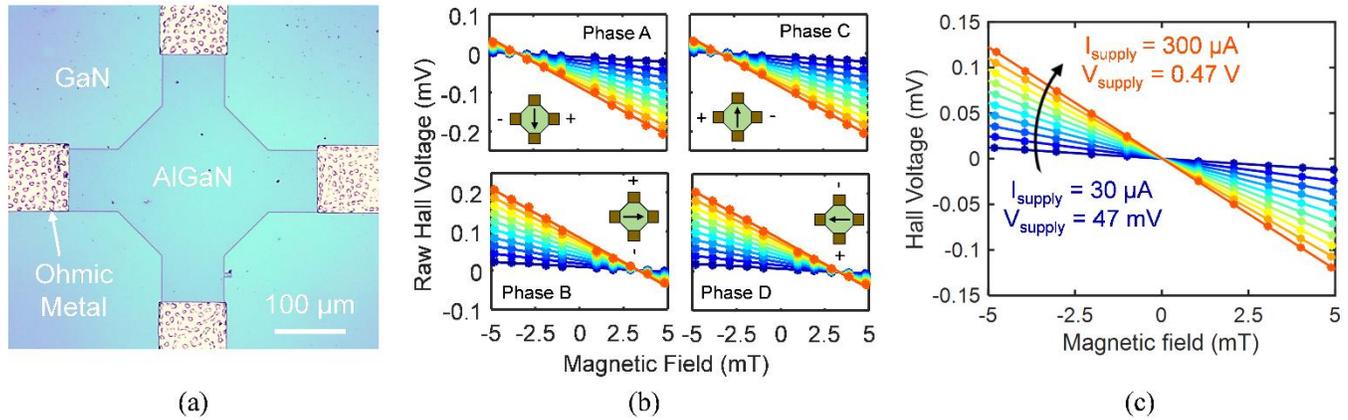

Fig. 1 (a) Optical image of the AlGaN/GaN 2DEG Hall-effect plate. b) Raw Hall voltage output from various magnetic fields under different supply current conditions for the 4 different measurement phases. Insets show measurement configuration used in each phase of current spinning: the arrows indicate supply current direction, and "+" and "-" indicate voltage measurement terminals. (c) Sensed Hall voltage (via current-spinning) under varied magnetic field and supply current conditions from 30 to 300 µA in increments of 30 µA. Fig. 1b uses the same supply conditions.

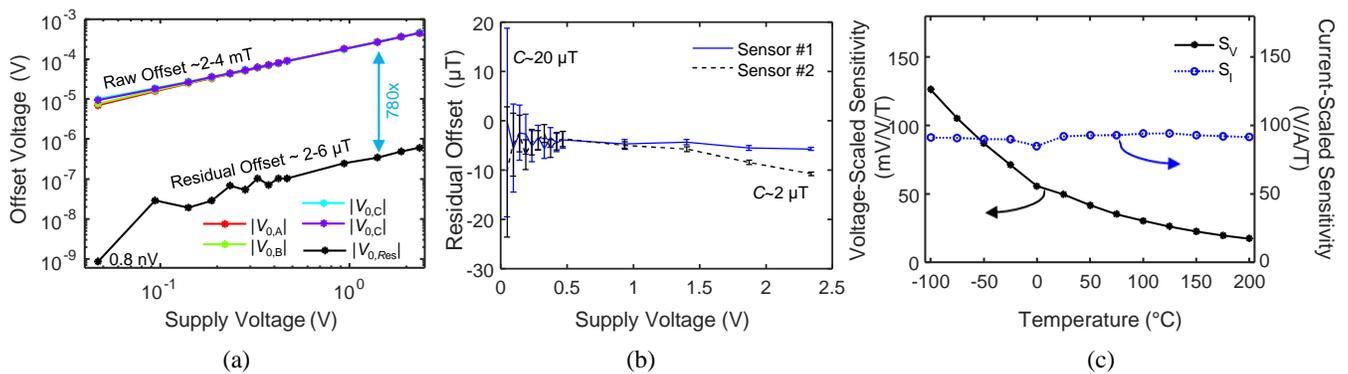

Fig. 2 (a) Magnitude of average offset voltages (raw and residual) measured in near-zero field for 300 measurements/datum. (b) Magnetic residual offset with respect to supply voltage, with 1-σ confidence ($C$) scale bars. (c) Sensor sensitivity scaled with voltage ($S_V$) and current ($S_I$), across various temperatures. $S_V$ decreases rapidly with temperature, and $S_I$ remains fairly constant.